**High-resolution path-integral development of financial options**

Lester Ingber

Lester Ingber Research

POB 06440 Sears Tower, Chicago, IL 60606

and

DRW Investments LLC

311 S Wacker Dr, Ste 900, Chicago, IL 60606

ingber@ingber.com, ingber@alumni.caltech.edu

## ABSTRACT

The Black-Scholes theory of option pricing has been considered for many years as an important but very approximate zeroth-order description of actual market behavior. We generalize the functional form of the diffusion of these systems and also consider multi-factor models including stochastic volatility. Daily Eurodollar futures prices and implied volatilities are fit to determine exponents of functional behavior of diffusions using methods of global optimization, Adaptive Simulated Annealing (ASA), to generate tight fits across moving time windows of Eurodollar contracts. These short-time fitted distributions are then developed into long-time distributions using a robust non-Monte Carlo path-integral algorithm, PATHINT, to generate prices and derivatives commonly used by option traders.

Keywords: options; eurodollar; volatility; path integral; optimization; statistical mechanics

# 1. INTRODUCTION

## 1.1. Background

There always is much interest in developing more sophisticated pricing models for financial instruments. In particular, there currently is much interest in improving option pricing models, particularly with respect to stochastic variables [1-4].

The standard Black-Scholes (BS) theory assumes a lognormal distribution of market prices, i.e., a diffusion linearly proportional to the market price. However, many texts include outlines of more general diffusions proportional to an arbitrary power of the market price [5].

The above aspects of stochastic volatility and of more general functional dependencies of diffusions are most often "swept under the rug" of a simple lognormal form. Experienced traders often use their own intuition to put volatility "smiles" into the BS theoretical constant coefficient in the BS lognormal distribution to compensate for these aspects.

It is generally acknowledged that since the market crash of 1987, markets have been increasingly difficult to describe using the BS model, and so better modelling and computational techniques should be used traders [6], although in practice simple BS models are the rule rather than the exception simply because they are easy to use [7]. To a large extent, previous modelling that has included stochastic volatility and multiple factors has been driven more by the desire to either delve into mathematics tangential to these issues, or to deal only with models that can accommodate closed-form algebraic expressions. We do not see much of the philosophy in the literature that has long driven the natural sciences: to respect first raw data, secondly models of raw data, and finally the use of numerical techniques that do not excessively distort models for the sake of ease of analysis and speed of computation. Indeed, very often the reverse set of priorities is seen in mathematical finance.

## 1.2. Our Approach

We have addressed the above issues in detail within the framework of a previously developed statistical mechanics of financial markets (SMFM) [8-13].

Our approach requires three sensible parts. Part one is the formulation of the model, which to some extent also involves specification of the specific market(s) data to be addressed. Part two is the fitting of the model to specific market data. Part three is the use of the resulting model to calculate option prices and their Greeks (partial derivatives of the prices with respect to their independent variables), which are used as risk parameters by traders. Each part requires some specific numerical tuning to the market under consideration.

The first part was to develop the algebraic model to replace/generalize BS, including the possibility of also addressing how to handle data regions not previously observed in trading. This is not absurd; current BS models perform integrals that must include a much influence from fat tails that include data regions never seen or likely to be seen in real-world markets. There are some issues as to whether we should take seriously the notion that the market is strongly driven by some element of a "self-fulfilling prophesy" by the BS model [14], but in any case our models have parameters to handle a wide range of possible cases that might arise.

We have developed two parallel tracks starting with part one, a one-factor and a two-factor model. The two-factor model includes stochastic volatility. At first we sensed the need to develop this two-factor model, and we now see that this is at the least an important benchmark against which to judge the worth of the one-factor model.

The second part was to fit the actual raw data so we can come up with real distributions. Some tests illustrated that standard quasi-linear fitting routines, could not get the proper fits, and so we used a more powerful global optimization, Adaptive Simulated Annealing (ASA) [15]. Tuning and selection of the time periods to perform the fits to the data were not trivial aspects of this research. Practical decisions had to be made on the time span of data to be fit and how to aggregate the fits to get sensible "fair values" for reasonable standard deviations of the exponents in the diffusions.

The third part was to develop Greeks and risk parameters from these distributions without making premature approximations just to ease the analysis. Perhaps someday, simple approximations and intuitions similar to what traders now use for BS models will be available for these models, but we do not think the best approach is to start out with such approximations until we first see proper calculations, especially in this uncharted territory. When it seemed that Cox-Ross-Rubenstein (CRR) standard tree codes (discretized approximations to partial differential equations) [16] were not stable for general exponents, i.e., for other than the lognormal case, we turned to a PATHINT code developed a decade ago for some hard nonlinear multifactor problems [17], e.g., combat analyses [18], neuroscience [19,20], and potentially chaotic systems [21,22]. In 1990 and 1991 papers on financial applications, it was mentioned how these techniques could be used for stochastic interest rates and bonds [9,10]. The modifications required here for one-factor European and then American cases went surprisingly smoothly; we still had to tune the meshes, etc. The two-factor model presented a technical problem to the algorithm, which we have reasonably handled using a combination of selection of the model in part one and a reasonable approach to developing the meshes.

### 1.3. Outline of Paper

Section 1 is this introduction. Section 2 describes the nature of Eurodollar (ED) futures data and the evidence for stochastic volatility. Section 3 outlines the algebra of modelling options, including the standard BS theory and our generalizations. Section 4 outlines the three equivalent mathematical representations used by SMFM; this is required to understand the development of the short-time distribution that defines the cost function we derive for global optimization, as well as the numerical methods we have developed to calculate the long-time evolution of these short-time distributions. Section 5 outlines ASA and explains its use to fit short-time probability distributions defined by our models to the Eurodollar data; we offer the fitted exponent in the diffusion as a new important technical indicator of market behavior. Section 6 outlines PATHINT and explains its use to develop long-time probability distributions from the fitted short-time probability distributions, for both the one-factor and two-factor tracks. Section 7 describes how we use these long-time probability distributions to calculate European and American option prices and Greeks; here we give numerical tests of our approach to BS CRR algorithms. Section 8 is our conclusion.

## 2. DATA

### 2.1. Eurodollars

Eurodollars are fixed-rate time deposits held primarily by overseas banks, but denominated in US dollars. They are not subject to US banking regulations and therefore tend to have a tighter bid-ask spread than deposits held in the United States [23].

## 2.2. Futures

The three-month Eurodollar futures contract is one of the most actively traded futures markets in the world. The contract is quoted as an index where the yield is equal to the Eurodollar price subtracted from 100. This yield is equal to the fixed rate of interest paid by Eurodollar time deposits upon maturity and is expressed as an annualized interest rate based on a 360-day year. The Eurodollar futures are cash settled based on the 90-day London Interbank Offer Rate (LIBOR). A "notional" principal amount of $1 million, is used to determine the change in the total interest payable on a hypothetical underlying time deposit, but is never actually paid or received [23].

Currently a total of 40 quarterly Eurodollar futures contracts (or ten years worth) are listed, with expirations annually in March, June, September and December.

## 2.3. Options on Futures

The options traded on the Eurodollar futures include not only 18 months of options expiring at the same time as the underlying future, but also various short dated options which themselves expire up to one year prior to the expiration of the underlying futures contract.

## 2.4. Front/Back Month Contracts

For purposes of risk minimization, as discussed in a previous paper [4], traders put on spreads across a variety of option contracts. One common example is to trade the spread on contracts expiring one year apart, where the future closer to expiration is referred to as the front month contract, and the future expiring one year later is called the back month. The availability of short dated or "mid-curve" options which are based on an underlying back month futures contract, but expire at the same time as the front month, allow one to trade the volatility ratios of the front and back month futures contracts without having to take the time differences in option expirations into consideration. We studied the volatilities of these types of front and back month contracts. Here, we give analyses with respect only to quarterly data longer than six months from expiration.

## 2.5. Stochastic Volatility

Below we develop two-factor models to address stochastic volatility. In a previous paper, we have performed empirical studies of Eurodollar futures to support the necessity of dealing with these issues [4].

## 3. MODELS

### 3.1. Random walk model

The use of Brownian motion as a model for financial systems is generally attributed to Bachelier [24], though he incorrectly intuited that the noise scaled linearly instead of as the square root relative to the random log-price variable. Einstein is generally credited with using the correct mathematical description in a larger physical context of statistical systems. However, several studies imply that changing prices of many markets do not follow a random walk, that they may have long-term dependences in price correlations, and that they may not be efficient in quickly arbitraging new information [25-27]. A random walk for returns, rate of change of prices over prices, is described by a Langevin equation with simple additive noise $\eta$, typically representing the continual random influx of information into the market.

$$\dot{\Gamma} = -\gamma_1 + \gamma_2 \eta \ ,$$

$$\dot{\Gamma} = d\Gamma/dt \ ,$$

$$< \eta(t) >_\eta = 0 \ , \ < \eta(t), \eta(t') >_\eta = \delta(t - t') \ , \tag{1}$$

where $\gamma_1$ and $\gamma_2$ are constants, and $\Gamma$ is the logarithm of (scaled) price. Price, although the most dramatic observable, may not be the only appropriate dependent variable or order parameter for the system of markets [28]. This possibility has also been called the "semistrong form of the efficient market hypothesis" [25].

The generalization of this approach to include multivariate nonlinear nonequilibrium markets led to a model of statistical mechanics of financial markets (SMFM) [8].

### 3.2. Black-Scholes (BS) Theory

The standard partial-differential equation used to formulate most variants of Black-Scholes (BS) models describing the market value of an option, $V$, is

$$\frac{\partial V}{\partial t} + \frac{1}{2} \sigma^2 S^2 \frac{\partial^2 V}{\partial S^2} + rS \frac{\partial V}{\partial S} - rV = 0 , \tag{2}$$

where $S$ is the asset price, and $\sigma$ is the standard deviation, or volatility of $S$, and $r$ is the short-term interest rate. The solution depends on boundary conditions, subject to a number of interpretations, some requiring minor transformations of the basic BS equation or its solution. For example, the basic equation can apply to a number of one-dimensional models of interpretations of prices given to $V$, e.g., puts or calls, and to $S$, e.g., stocks or futures, dividends, etc.

For instance, if $V$ is set to $C$, a call on an European option with exercise price $X$ with maturity at $T$, the solution is

$$C(S, t) = SN(d_1) - Xe^{-r(T-t)} N(d_2) ,$$

$$d_1 = \frac{\ln(S/X) + (r + \frac{1}{2} \sigma^2)(T-t)}{\sigma(T-t)^{1/2}} ,$$

$$d_2 = \frac{\ln(S/X) + (r - \frac{1}{2} \sigma^2)(T-t)}{\sigma(T-t)^{1/2}} . \tag{3}$$

In practice, the volatility $\sigma$ is the least known parameter in this equation, and its estimation is generally the most important part of pricing options. Usually the volatility is given in a yearly basis, baselined to some standard, e.g., 252 trading days per year, or 360 or 365 calendar days. Therefore, all values of volatility given in the graphs in this paper, based on daily data, would be annualized by multiplying the standard deviations of the yields by $\sqrt{252} = 15.87$. We have used this factor to present our implied volatilities as daily movements.

### 3.3. Some Key Issues in Derivation of BS

The basic BS model considers a portfolio in terms of **delta** ($\Delta$),

$$\Pi = V - \Delta S , \tag{4}$$

in a market with Gaussian-Markovian ("white") noise $X$ and drift $\mu$,

$$\frac{dS}{S} = \sigma dX + \mu dt \ , \tag{5}$$

where $V(S, t)$ inherits a random process from $S$,

$$dV = \sigma S \frac{\partial V}{\partial S} dX + \left( \mu S \frac{\partial V}{\partial S} + \frac{1}{2} \sigma^2 S^2 \frac{\partial^2 V}{\partial S^2} + \frac{\partial V}{\partial t} \right) dt \ . \tag{6}$$

This yields

$$d\Pi = \sigma \left( \frac{\partial V}{\partial S} - \Delta \right) dX + \left( \mu S \frac{\partial V}{\partial S} + \frac{1}{2} \sigma^2 S^2 \frac{\partial^2 V}{\partial S^2} + \frac{\partial V}{\partial t} - \mu \Delta S \right) dt \ . \tag{7}$$

The expected risk-neutral return of $\Pi$ is

$$d\Pi = r \Pi dt = r(V - \Delta S) dt \ . \tag{8}$$

Options $V$ on futures $F$ can be derived, e.g., using simple transformations to take cost of carry into consideration, such as

$$F = Se^{r(T-t)} \ , \tag{9}$$

and setting

$$d\Pi = rV \ dt \ . \tag{10}$$

The corresponding BS equation for futures $F$ is

$$\frac{\partial V}{\partial t} + \frac{1}{2} \sigma^2 F^2 \frac{\partial^2 V}{\partial S^2} - rV = 0 \ . \tag{11}$$

At least two advantages are present if $\Delta$ is chosen such that

$$\Delta = \frac{\partial V}{\partial S} \ . \tag{12}$$

Then, the portfolio can be instantaneously "risk-neutral," in terms of zeroing the coefficient of $X$, as well as independent of the direction of market, in terms of zeroing the coefficient of $\mu$. For the above example of $V = C$,

$$\Delta = N(d_1) \ . \tag{13}$$

Other trading strategies based on this simple model use similar constructs as risk parameters, e.g., **gamma** ($\Gamma$), **theta** ($\Theta$), **vega** ($\Upsilon$), **rho** ($\rho$) [5],

$$\Gamma = \frac{\partial^2 \Pi}{\partial S^2} \ ,$$

$$\Theta = \frac{\partial \Pi}{\partial t} \ ,$$

$$\Upsilon = \frac{\partial \Pi}{\partial \sigma} \ ,$$

$$\rho = \frac{\partial \Pi}{\partial r} \ . \tag{14}$$

The BS equation, Eq. (2), may be written as

$$\Theta + rS\Delta + \frac{1}{2}(\sigma S)^2 \Gamma = rf \ . \tag{15}$$

### 3.4. $S^x$ Models

Our two-factor model includes stochastic volatility $\sigma$ of the underlying $S$,

$$dS = \mu \, dt + \sigma \, F(S, S_0, S_\infty, x, y) \, dz_S \ ,$$

$$d\sigma = \nu \, dt + \varepsilon \, dz_\sigma \ ,$$

$$< dz_i > = 0 \ , \ i = \{S, \sigma\} \ ,$$

$$< dz_i(t) \, dz_j(t') > = dt \, \delta(t - t') \ , \ i = j \ ,$$

$$< dz_i(t) \, dz_j(t') > = \rho \, dt \, \delta(t - t') \ , \ i \neq j \ ,$$

$$F(S, S_0, S_\infty, x, y) = \begin{cases} S, & S < S_0 \\ S^x S_0^{1-x}, & S_0 \leq S \leq S_\infty \\ S^y S_0^{1-x} S_\infty^{x-y}, & S > S_\infty \end{cases} \ , \tag{16}$$

where $S_0$ and $S_\infty$ are selected to lie outside the data region used to fit the other parameters, e.g., $S_0 = 1$ and $S_\infty = 20$ for fits to Eurodollar futures which historically have a very tight range relative to other markets. We have used the Black-Scholes form $F = S$ inside $S < S_0$ to obtain the usual benefits, e.g., no negative prices as the distribution is naturally excluded from $S < 0$ and preservation of put-call parity. Put-call parity for European options is derived quite independent of any mathematical model of options [5]. In its simplest form, it is given by

$$c + Xe^{-r(T-t)} = p + S \ , \qquad (17)$$

where $c$ ($p$) is the fair price of a call (put), $X$ is the strike price, $r$ is the risk-free interest rate, $t$ is the present time, $T$ is the time of expiration, and $S$ is the underlying market. We have taken $y = 0$, a normal distribution, to reflect total ignorance of markets outside the range of $S > S_\infty$. The one-factor model just assumes a constant $\sigma$. It is often noted that BS models incorrectly include untenable contributions from large $S$ regions because of their fat tails [29]. (If we wished to handle negative interest rates, ED prices > 100, we would move shift the $S = 0$ axis to some $S < 0$ value.)

We found that the abrupt, albeit continuous, changes across $S_0$ especially for $x \le 0$ did not cause any similar effects in the distributions evolved using these diffusions, as reported below.

The formula for pricing an option $P$, derived in a Black-Scholes generalized framework after factoring out interest-rate discounting, is equivalent to using the form

$$dS = \mu \, S \, dt + \sigma \, F(S, S_0, S_\infty, x, y) \, dz_S \ ,$$

$$d\sigma = \nu \, dt + \varepsilon \, dz_\sigma \ . \qquad (18)$$

We experimented with some alternative functional forms, primarily to apply some smooth cutoffs across the above three regions of $S$. For example, we used $F'$, a function $F$ designed to revert to the lognormal Black-Scholes model in several limits,

$$F'(S, S_0, S_\infty, x) = S \, C_0 + (1 - C_0) \, (S^x \, S_0^{1-x} \, C_\infty + S_0 (1 - C_\infty)) \ ,$$

$$C_0 = \exp \left[ -\left( \frac{S}{S_0} \, \frac{|1-x|}{1 + |1-x|} \right)^{2-x|+1} \right] \ ,$$

$$C_\infty = \exp \left[ -\left( \frac{S}{S_\infty} \right)^2 \right] \ ,$$

$$\lim_{S \to \infty, \, x \ne 1} F'(S, S_0, S_\infty, x) = S_0 = \text{constant} \ ,$$

$$\lim_{S \to 0^+} F'(S, S_0, S_\infty, x) = \lim_{x \to 1} F'(S, S_0, S_\infty, x) = S \ . \tag{19}$$

However, our fits were most sensitive to the data when we permitted the central region to be pure $S^x$ using $F$ above.

### 3.4.1. Various $F(S, x)$ Diffusions

Fig. 1 gives examples of $F(S, S_0, S_\infty, x, y) \, dz_S$ for $x$ in {-1, 0, 1, 2}. The other parameters are $S = 5$, $S_0 = 0.5$, $S_\infty = 20$, $y = 0$.

---

Fig. 1.

---

## 4. STATISTICAL MECHANICS OF FINANCIAL MARKETS (SMFM)

### 4.1. Statistical Mechanics of Large Systems

Aggregation problems in nonlinear nonequilibrium systems typically are "solved" (accommodated) by having new entities/languages developed at these disparate scales in order to efficiently pass information back and forth. This is quite different from the nature of quasi-equilibrium quasi-linear systems, where thermodynamic or cybernetic approaches are possible. These approaches typically fail for nonequilibrium nonlinear systems.

Many systems are aptly modeled in terms of multivariate differential rate-equations, known as Langevin equations,

$$\dot{M}^G = f^G + \hat{g}^G_j \eta^j \ , (G = 1, \cdots, \Lambda) \ , (j = 1, \cdots, N) \ ,$$

$$\dot{M}^G = dM^G / dt \ ,$$

$$< \eta^j(t) >_\eta = 0 \ , \ < \eta^j(t), \eta^{j'}(t') >_\eta = \delta^{jj'} \delta(t - t') \ , \tag{20}$$

where $f^G$ and $\hat{g}^G_j$ are generally nonlinear functions of mesoscopic order parameters $M^G$, $j$ is a microscopic index indicating the source of fluctuations, and $N \geq \Lambda$. The Einstein convention of summing over repeated indices is used. Vertical bars on an index, e.g., |j|, imply no sum is to be taken on repeated

indices.

Via a somewhat lengthy, albeit instructive calculation, outlined in several other papers [8,10,30], involving an intermediate derivation of a corresponding Fokker-Planck or Schrödinger-type equation for the conditional probability distribution $P[M(t)|M(t_0)]$, the Langevin rate Eq. (20) is developed into the more useful probability distribution for $M^G$ at long-time macroscopic time event $t = (u + 1)\theta + t_0$, in terms of a Stratonovich path-integral over mesoscopic Gaussian conditional probabilities [31-35]. Here, macroscopic variables are defined as the long-time limit of the evolving mesoscopic system.

The corresponding Schrödinger-type equation is [33,34]

$$\partial P/\partial t = \frac{1}{2} (g^{GG'} P)_{,GG'} - (g^G P)_{,G} + V \ ,$$

$$g^{GG'} = k_T \delta^{jk} \hat{g}^G_j \hat{g}^{G'}_k \ ,$$

$$g^G = f^G + \frac{1}{2} \delta^{jk} \hat{g}^{G'}_j \hat{g}^G_{k,G'} \ ,$$

$$[\cdots]_{,G} = \partial[\cdots]/\partial M^G \ . \tag{21}$$

This is properly referred to as a Fokker-Planck equation when $V \equiv 0$. Note that although the partial differential Eq. (21) contains information regarding $M^G$ as in the stochastic differential Eq. (20), all references to $j$ have been properly averaged over. I.e., $\hat{g}^G_j$ in Eq. (20) is an entity with parameters in both microscopic and mesoscopic spaces, but $M$ is a purely mesoscopic variable, and this is more clearly reflected in Eq. (21).

The path integral representation is given in terms of the "Feynman" Lagrangian $L$.

$$P[M_t|M_{t_0}]dM(t) = \int \cdots \int \mathcal{D}M \exp(-S)\delta[M(t_0) = M_0]\delta[M(t) = M_t] \ ,$$

$$S = k_T^{-1} \min \int_{t_0}^{t} dt' L \ ,$$

$$\mathcal{D}M = \lim_{u \to \infty} \prod_{\rho=1}^{u+1} g^{1/2} \prod_G (2\pi\theta)^{-1/2} dM^G_\rho \ ,$$

$$L(\dot{M}^G, M^G, t) = \frac{1}{2} (\dot{M}^G - h^G) g_{GG'} (\dot{M}^{G'} - h^{G'}) + \frac{1}{2} h^G_{;G} + R/6 - V \ ,$$

$$h^G = g^G - \frac{1}{2} g^{-1/2} (g^{1/2} g^{GG'})_{,G'} \, ,$$

$$g_{GG'} = (g^{GG'})^{-1} \, ,$$

$$g = \det(g_{GG'}) \, ,$$

$$h^G_{\;;G} = h^G_{,G} + \Gamma^F_{GF} h^G = g^{-1/2} (g^{1/2} h^G)_{,G} \, ,$$

$$\Gamma^F_{JK} \equiv g^{LF} [JK, L] = g^{LF} (g_{JL,K} + g_{KL,J} - g_{JK,L}) \, ,$$

$$R = g^{JL} R_{JL} = g^{JL} g^{JK} R_{FJKL} \, ,$$

$$R_{FJKL} = \frac{1}{2} (g_{FK,JL} - g_{JK,FL} - g_{FL,JK} + g_{JL,FK}) + g_{MN} (\Gamma^M_{FK} \Gamma^N_{JL} - \Gamma^M_{FL} \Gamma^N_{JK}) \, . \tag{22}$$

Mesoscopic variables have been defined as $M^G$ in the Langevin and Fokker-Planck representations, in terms of their development from the microscopic system labeled by $j$. The Riemannian curvature term $R$ arises from nonlinear $g_{GG'}$, which is a bona fide metric of this space [33]. Even if a stationary solution, i.e., $\dot{M}^G = 0$, is ultimately sought, a necessarily prior stochastic treatment of $\dot{M}^G$ terms gives rise to these Riemannian "corrections." Even for a constant metric, the term $h^G_{\;;G}$ contributes to $L$ for a nonlinear mean $h^G$. $V$ may include terms such as $\sum_{T'} J_{T'G} M^G$, where the Lagrange multipliers $J_{T'G}$ are constraints on $M^G$, which are advantageously modeled as extrinsic sources in this representation; they too may be time-dependent.

For our purposes, the above Feynman Lagrangian defines a kernel of the short-time conditional probability distribution, in the curved space defined by the metric, in the limit of continuous time, whose iteration yields the solution of the previous partial differential equation Eq. (21). This differs from the Lagrangian which satisfies the requirement that the action is stationary to the first order in $dt$ — the WKBJ approximation, but which does not include the first-order correction to the WKBJ approximation as does the Feynman Lagrangian. This latter Lagrangian differs from the Feynman Lagrangian, essentially by replacing $R/6$ above by $R/12$ [36]. In this sense, the WKBJ Lagrangian is more useful for some theoretical discussions [37]. However, the use of the Feynman Lagrangian coincides with the numerical method we present here using the PATHINT code.

Using the variational principle, $J_{TG}$ may also be used to constrain $M^G$ to regions where they are empirically bound. More complicated constraints may be affixed to $L$ using methods of optimal control theory [38]. With respect to a steady state $\bar{P}$, when it exists, the information gain in state $P$ is defined by

$$\Upsilon[P] = \int \cdots \int \underset{\sim}{D}M' \, P \ln (P/\bar{P}) \, ,$$

$$\underset{\sim}{D}M' = \underset{\sim}{D}M/dM_{u+1} \, . \tag{23}$$

In the economics literature, there appears to be sentiment to define Eq. (20) by the Ito, rather than the Stratonovich prescription. It is true that Ito integrals have Martingale properties not possessed by Stratonovich integrals [39] which leads to risk-neural theorems for markets [40,41], but the nature of the proper mathematics should eventually be determined by proper aggregation of relatively microscopic models of markets. It should be noted that virtually all investigations of other physical systems, which are also continuous time models of discrete processes, conclude that the Stratonovich interpretation coincides with reality, when multiplicative noise with zero correlation time, modeled in terms of white noise $\eta^j$, is properly considered as the limit of real noise with finite correlation time [42]. The path integral succinctly demonstrates the difference between the two: The Ito prescription corresponds to the prepoint discretization of $L$, wherein $\theta \dot{M}(t) \to M_{\rho+1} - M_\rho$ and $M(t) \to M_\rho$. The Stratonovich prescription corresponds to the midpoint discretization of $L$, wherein $\theta \dot{M}(t) \to M_{\rho+1} - M_\rho$ and $M(t) \to \frac{1}{2}(M_{\rho+1} + M_\rho)$. In terms of the functions appearing in the Fokker-Planck Eq. (21), the Ito prescription of the prepoint discretized Lagrangian, $L_I$, is relatively simple, albeit deceptively so because of its nonstandard calculus.

$$L_I(\dot{M}^G, M^G, t) = \frac{1}{2}(\dot{M}^G - g^G)g_{GG'}(\dot{M}^{G'} - g^{G'}) - V \, . \tag{24}$$

In the absence of a nonphenomenological microscopic theory, the difference between a Ito prescription and a Stratonovich prescription is simply a transformed drift [36].

There are several other advantages to Eq. (22) over Eq. (20). Extrema and most probable states of $M^G$, $\ll M^G \gg$, are simply derived by a variational principle, similar to conditions sought in previous studies [43]. In the Stratonovich prescription, necessary, albeit not sufficient, conditions are given by

$$\delta_G L = L_{,G} - L_{,\dot{G}:t} = 0 \, ,$$

$$L_{,\dot{G}:t} = L_{,\dot{G}G'}\dot{M}^{G'} + L_{,\dot{G}\dot{G}'}\ddot{M}^{G'} .$$ (25)

For stationary states, $\dot{M}^G = 0$, and $\partial \bar{L}/\partial \bar{M}^G = 0$ defines $\ll \bar{M}^G \gg$, where the bars identify stationary variables; in this case, the macroscopic variables are equal to their mesoscopic counterparts. [Note that $\bar{L}$ is *not* the stationary solution of the system, e.g., to Eq. (21) with $\partial P/\partial t = 0$. However, in some cases [44], $\bar{L}$ is a definite aid to finding such stationary states.] Many times only properties of stationary states are examined, but here a temporal dependence is included. E.g., the $\dot{M}^G$ terms in $L$ permit steady states and their fluctuations to be investigated in a nonequilibrium context. Note that Eq. (25) must be derived from the path integral, Eq. (22), which is at least one reason to justify its development.

## 4.2. Correlations

Correlations between variables are modeled explicitly in the Lagrangian as a parameter usually designated $\rho$ (not to be confused with the Rho Greek calculated for options). This section uses a simple two-factor model to develop the correspondence between the correlation $\rho$ in the Lagrangian and that among the commonly written Weiner distributions $dz$.

Consider coupled stochastic differential equations

$$dr = f^r(r, l)dt + \hat{g}^r(r, l)\sigma_1 dz_1 ,$$

$$dl = f^l(r, l)dt + \hat{g}^l(r, l)\sigma_2 dz_2 ,$$

$$< dz_i >= 0 , i = \{1, 2\} ,$$

$$< dz_i(t)dz_j(t') >= dt\delta(t - t') , i = j ,$$

$$< dz_i(t)dz_j(t') >= \rho dt\delta(t - t') , i \neq j ,$$

$$\delta(t - t') = \begin{cases} 0 , , & t \neq t' , \\ 1 , & t = t' , \end{cases}$$ (26)

where $< . >$ denotes expectations.

These can be rewritten as Langevin equations (in the Itô prepoint discretization)

$$dr/dt = f^r + \hat{g}^r \sigma_1 (\gamma^+ n_1 + \text{sgn}\rho \, \gamma^- n_2) ,$$

$$dl/dt = g^l + \hat{g}^l \sigma_2 (\text{sgn}\, \rho\, \gamma^- n_1 + \gamma^+ n_2) \ ,$$

$$\gamma^\pm = \frac{1}{\sqrt{2}} \, [1 \pm (1 - \rho^2)^{1/2}]^{1/2} \ ,$$

$$n_i = (dt)^{1/2} p_i \ , \tag{27}$$

where $p_1$ and $p_2$ are independent [0,1] Gaussian distributions.

The equivalent short-time probability distribution, $P$, for the above set of equations is

$$P = g^{1/2} (2\pi dt)^{-1/2} \exp(-L dt) \ ,$$

$$L = \frac{1}{2} \, F^\dagger \underline{g} F \ ,$$

$$F = \begin{pmatrix} dr/dt - f^r \\ dl/dt - f^l \end{pmatrix} ,$$

$$g = \det(\underline{g}) \ ,$$

$$k = 1 - \rho^2 \ . \tag{28}$$

$\underline{g}$, the metric in $\{r, l\}$-space, is the inverse of the covariance matrix,

$$\underline{g}^{-1} = \begin{pmatrix} (\hat{g}^r \sigma_1)^2 & \rho \hat{g}^r \hat{g}^l \sigma_1 \sigma_2 \\ \rho \hat{g}^r \hat{g}^l \sigma_1 \sigma_2 & (\hat{g}^l \sigma_2)^2 \end{pmatrix} . \tag{29}$$

The above also corrects previous papers which inadvertently dropped the sgn factors in the above [9,10,17].

## 5.  ADAPTIVE SIMULATED ANNEALING (ASA) FITS

### 5.1.  ASA Outline

The algorithm developed which is now called Adaptive Simulated Annealing (ASA) [45] fits short-time probability distributions to observed data, using a maximum likelihood technique on the Lagrangian. This algorithm has been developed to fit observed data to a theoretical cost function over a $D$-dimensional parameter space [45], adapting for varying sensitivities of parameters during the fit.  The ASA code can be obtained at no charge, via WWW from http://www.ingber.com/ or via FTP from ftp.ingber.com [15].

### 5.1.1. General description

Simulated annealing (SA) was developed in 1983 to deal with highly nonlinear problems [46], as an extension of a Monte-Carlo importance-sampling technique developed in 1953 for chemical physics problems. It helps to visualize the problems presented by such complex systems as a geographical terrain. For example, consider a mountain range, with two "parameters," e.g., along the North–South and East–West directions. We wish to find the lowest valley in this terrain. SA approaches this problem similar to using a bouncing ball that can bounce over mountains from valley to valley. We start at a high "temperature," where the temperature is an SA parameter that mimics the effect of a fast moving particle in a hot object like a hot molten metal, thereby permitting the ball to make very high bounces and being able to bounce over any mountain to access any valley, given enough bounces. As the temperature is made relatively colder, the ball cannot bounce so high, and it also can settle to become trapped in relatively smaller ranges of valleys.

We imagine that our mountain range is aptly described by a "cost function." We define probability distributions of the two directional parameters, called generating distributions since they generate possible valleys or states we are to explore. We define another distribution, called the acceptance distribution, which depends on the difference of cost functions of the present generated valley we are to explore and the last saved lowest valley. The acceptance distribution decides probabilistically whether to stay in a new lower valley or to bounce out of it. All the generating and acceptance distributions depend on temperatures.

In 1984 [47], it was established that SA possessed a proof that, by carefully controlling the rates of cooling of temperatures, it could statistically find the best minimum, e.g., the lowest valley of our example above. This was good news for people trying to solve hard problems which could not be solved by other algorithms. The bad news was that the guarantee was only good if they were willing to run SA forever. In 1987, a method of fast annealing (FA) was developed [48], which permitted lowering the temperature exponentially faster, thereby statistically guaranteeing that the minimum could be found in some finite time. However, that time still could be quite long. Shortly thereafter, Very Fast Simulated Reannealing (VFSR) was developed in 1987 [45], now called Adaptive Simulated Annealing (ASA), which is exponentially faster than FA.

ASA has been applied to many problems by many people in many disciplines [49-51]. The feedback of many users regularly scrutinizing the source code ensures its soundness as it becomes more flexible and powerful.

### 5.1.2. Mathematical outline

ASA considers a parameter $\alpha_k^i$ in dimension $i$ generated at annealing-time $k$ with the range

$$\alpha_k^i \in [A_i, B_i] , \tag{30}$$

calculated with the random variable $y^i$,

$$\alpha_{k+1}^i = \alpha_k^i + y^i(B_i - A_i) ,$$

$$y^i \in [-1, 1] . \tag{31}$$

The generating function $g_T(y)$ is defined,

$$g_T(y) = \prod_{i=1}^{D} \frac{1}{2(|y^i| + T_i)\ln(1 + 1/T_i)} \equiv \prod_{i=1}^{D} g_T^i(y^i) , \tag{32}$$

where the subscript $i$ on $T_i$ specifies the parameter index, and the $k$-dependence in $T_i(k)$ for the annealing schedule has been dropped for brevity. Its cumulative probability distribution is

$$G_T(y) = \int_{-1}^{y^1} \cdots \int_{-1}^{y^D} dy'^1 \cdots dy'^D \, g_T(y') \equiv \prod_{i=1}^{D} G_T^i(y^i) ,$$

$$G_T^i(y^i) = \frac{1}{2} + \frac{\text{sgn}\,(y^i)}{2} \, \frac{\ln(1 + |y^i|/T_i)}{\ln(1 + 1/T_i)} . \tag{33}$$

$y^i$ is generated from a $u^i$ from the uniform distribution

$$u^i \in U[0, 1] ,$$

$$y^i = \text{sgn}\,(u^i - \frac{1}{2})T_i[(1 + 1/T_i)^{|2u^i - 1|} - 1] . \tag{34}$$

It is straightforward to calculate that for an annealing schedule for $T_i$

$$T_i(k) = T_{0i} \exp(-c_i k^{1/D}) , \tag{35}$$

a global minima statistically can be obtained.  I.e.,

$$\sum_{k_0}^{\infty} g_k \approx \sum_{k_0}^{\infty} [\prod_{i=1}^{D} \frac{1}{2|y^i|c_i}] \frac{1}{k} = \infty \ . \tag{36}$$

Control can be taken over $c_i$, such that

$$T_{fi} = T_{0i} \exp(-m_i) \text{ when } k_f = \exp n_i \ ,$$

$$c_i = m_i \exp(-n_i/D) \ , \tag{37}$$

where $m_i$ and $n_i$ can be considered "free" parameters to help tune ASA for specific problems.

ASA has over 100 OPTIONS available for tuning.  A few important ones were used in this project.

### 5.1.3.  Reannealing

Whenever doing a multi-dimensional search in the course of a complex nonlinear physical problem, inevitably one must deal with different changing sensitivities of the $\alpha^i$ in the search.  At any given annealing-time, the range over which the relatively insensitive parameters are being searched can be "stretched out" relative to the ranges of the more sensitive parameters.  This can be accomplished by periodically rescaling the annealing-time $k$, essentially reannealing, every hundred or so acceptance-events (or at some user-defined modulus of the number of accepted or generated states), in terms of the sensitivities $s_i$ calculated at the most current minimum value of the cost function, $C$,

$$s_i = \partial C / \partial \alpha^i \ . \tag{38}$$

In terms of the largest $s_i = s_{\max}$, a default rescaling is performed for each $k_i$ of each parameter dimension, whereby a new index $k'_i$ is calculated from each $k_i$,

$$k_i \rightarrow k'_i \ ,$$

$$T'_{ik'} = T_{ik}(s_{\max}/s_i) \ ,$$

$$k'_i = (\ln(T_{i0}/T_{ik'})/c_i)^D \ . \tag{39}$$

$T_{i0}$ is set to unity to begin the search, which is ample to span each parameter dimension.

### 5.1.4.  Quenching

Another adaptive feature of ASA is its ability to perform quenching in a methodical fashion.  This is applied by noting that the temperature schedule above can be redefined as

$$T_i(k_i) = T_{0i} \exp(-c_i k_i^{Q_i/D}) \ ,$$

$$c_i = m_i \exp(-n_i Q_i/D) \ , \tag{40}$$

in terms of the "quenching factor" $Q_i$.  The sampling proof fails if $Q_i > 1$ as

$$\sum_k \prod_k^D 1/k^{Q_i/D} = \sum_k 1/k^{Q_i} < \infty \ . \tag{41}$$

This simple calculation shows how the "curse of dimensionality" arises, and also gives a possible way of living with this disease.  In ASA, the influence of large dimensions becomes clearly focussed on the exponential of the power of $k$ being $1/D$, as the annealing required to properly sample the space becomes prohibitively slow.  So, if resources cannot be committed to properly sample the space, then for some systems perhaps the next best procedure may be to turn on quenching, whereby $Q_i$ can become on the order of the size of number of dimensions.

The scale of the power of $1/D$ temperature schedule used for the acceptance function can be altered in a similar fashion.  However, this does not affect the annealing proof of ASA, and so this may used without damaging the sampling property.

### 5.2.  *x*-Indicator of Market Contexts

Our studies of contexts of markets well recognized by option traders to have significantly different volatility behavior show that the exponents $x$ are reasonably faithful indicators defining these different contexts.

We feel the two-factor model is more accurate because the data indeed demonstrate stochastic volatility [4].  We also note that the two-factor $x$'s are quite robust and uniform when being fit by ASA across the last few years.  This is not true of the one-factor ASA fitted $x$'s unless we do not use the Black-Scholes $\sigma$ as a parameter, but rather calculate as historical volatility during all runs.  Some results of two-factor studies and one-factor studies using a Black-Scholes $\sigma$ have been reported elsewhere [13].

Since $\sigma$ is not widely traded and arbitraged, to fit the two-factor model, we calculate this quantity as an historical volatility for both its prepoint and postpoint values. Some previous studies used a scaled implied volatility (which is calculated from a BS model). We use a standard deviation $\sigma'$,

$$\sigma' = \text{StdDev}\big(dS \, / \, F(S, S_0, S_\infty, x, y)\big) \, . \tag{42}$$

In the one-factor model, it does not make good numerical sense to have two free parameters in one term, i.e., $\sigma$ and $x$, as these cannot be fit very well within the variance the data. Instead, one method is to take guidance from the two-factor results, to set a scale for an effective $\sigma$, and then fit the parameter $x$. Another method it apply the above StdDev as a proxy for $\sigma$. Some motivation for this approach is given by considering collapsing a two-factor stochastic volatility model in one-factor model: The one-factor model now has an integral over the stochastic process in its diffusion term. The is integral is what we are approximating by using the standard deviation of a moving window of the data.

## 6. PATH-INTEGRAL (PATHINT) DEVELOPMENT

### 6.1. PATHINT Outline

The fits described above clearly demonstrate the need to incorporate stochastic volatility in option pricing models. If one-factor fits are desired, e.g., for efficiency of calculation, then at the least the exponent of price $x$ should be permitted to freely adapt to the data. In either case, it is required to develop a full set of Greeks for trading. To meet these needs, we have used a path-integral code, PATHINT, described below, with great success. At this time, the two-factor code takes too long to run for daily use, but it proves to be a good weekly baseline for the one-factor code.

The PATHINT algorithm develops the long-time probability distribution from the Lagrangian fit by the first optimization code. A robust and accurate histogram-based (non-Monte Carlo) path-integral algorithm to calculate the long-time probability distribution has been developed to handle nonlinear Lagrangians [18-20,22,52-54],

The histogram procedure recognizes that the distribution can be numerically approximated to a high degree of accuracy as sum of rectangles at points $M_i$ of height $P_i$ and width $\Delta M_i$. For convenience, just consider a one-dimensional system. The above path-integral representation can be rewritten, for each of its intermediate integrals, as

$$P(M; t + \Delta t) = \int dM' [g_s^{1/2} (2\pi \Delta t)^{-1/2} \exp(-L_s \Delta t)] P(M'; t) = \int dM' G(M, M'; \Delta t) P(M'; t) ,$$

$$P(M; t) = \sum_{i=1}^{N} \pi(M - M_i) P_i(t) ,$$

$$\pi(M - M_i) = \begin{cases} 0 , (M_i - \frac{1}{2} \Delta M_{i-1}) \leq M \leq (M_i + \frac{1}{2} \Delta M_i) , \\ 1 , \text{ otherwise} , \end{cases} \tag{43}$$

which yields

$$P_i(t + \Delta t) = T_{ij}(\Delta t) P_j(t) ,$$

$$T_{ij}(\Delta t) = \frac{2}{\Delta M_{i-1} + \Delta M_i} \int_{M_i - \Delta M_{i-1}/2}^{M_i + \Delta M_i/2} dM \int_{M_j - \Delta M_{j-1}/2}^{M_j + \Delta M_j/2} dM' G(M, M'; \Delta t) . \tag{44}$$

$T_{ij}$ is a banded matrix representing the Gaussian nature of the short-time probability centered about the (varying) drift.

Care must be used in developing the mesh in $\Delta M^G$, which is strongly dependent on the diagonal elements of the diffusion matrix, e.g.,

$$\Delta M^G \approx (\Delta t g^{|G||G|})^{1/2} . \tag{45}$$

Presently, this constrains the dependence of the covariance of each variable to be a nonlinear function of that variable, albeit arbitrarily nonlinear, in order to present a straightforward rectangular underlying mesh. Below we address how we have handled this problem in our two-factor stochastic-volatility model.

Fitting data with the short-time probability distribution, effectively using an integral over this epoch, permits the use of coarser meshes than the corresponding stochastic differential equation. The coarser resolution is appropriate, typically required, for numerical solution of the time-dependent path-integral: By considering the contributions to the first and second moments of $\Delta M^G$ for small time slices $\theta$, conditions on the time and variable meshes can be derived [52]. The time slice essentially is determined by $\theta \leq \bar{L}^{-1}$, where $\bar{L}$ is the "static" Lagrangian with $dM^G/dt = 0$, throughout the ranges of $M^G$ giving the most important contributions to the probability distribution $P$. The variable mesh, a function of $M^G$, is optimally chosen such that $\Delta M^G$ is measured by the covariance $g^{GG'}$, or $\Delta M^G \approx (g^{GG} \theta)^{1/2}$.

If the histogram method is further developed into a trapezoidal integration, then more accuracy can be expected at more complex boundaries [53]. Such problems does not arise here, and 6–7 significant figure accuracy is easily achieved provided great care is taken to develop the mesh as described above. For example, after setting the initial-condition discretized delta function *MSUPG* at the prepoint of an interval, the mesh going forward in $M^G$ is simply calculated stepwise using

$$\Delta M^G = (g^{GG}\theta)^{1/2} \ . \tag{46}$$

However, going backwards in $M^G$, an iterative procedure was used at each step, starting with an estimate from the prepoint and going forward again, until there was no mismatch. That much care is required for the mesh was observed in the original Wehner-Wolfer paper [52].

It is important to stress that very good numerical accuracy is required to get very good Greeks required for real-world trading. Many authors develop very efficient numerical schemes to get reasonable prices to 2 or 3 significant figures, but these methods often are not very good to enough significant figures to get good precision for the Greeks. Typical Monte Carlo methods are notorious for giving such poor results after very long computer runs. In particular, we do not believe that good Greeks required for trading can be obtained by using meshes obtained by other simpler algorithms [55].

The PATHINT algorithm in its present form can "theoretically" handle any n-factor model subject to its diffusion-mesh constraints. In practice, the calculation of 3-factor and 4-factor models likely will wait until giga-hertz speeds and giga-byte RAM are commonplace.

### 6.2. Development of Long-Time Probabilities

The noise determined empirically as the diffusion of the data is the same, independent of $x$ within our approach. Therefore, we scale different exponents such that the the diffusions, the square of the "basis-point volatilities" (BPV), are scaled to be equivalent. Then, there is not a very drastic change in option prices for different exponents $x$ for the strike $X$ set to the $S$ underlying, the at-the-money (ATM) strike. This is not the case for out of the money (OTM) or in the money (ITM) strikes, e.g., when exercising the strike would generate loss or profit, resp. This implies that current pricing models are not radically mispricing the markets, but there still are significant changes in Greeks using more sophisticated models.

### 6.3. Dependence of Probabilities on *S* and *x*

Fig. 2 gives examples of the short-time distribution evolved out to $T = 0.5$ year for $x$ in {-1, 0, 1, 2}, with 500 intermediate epochs/foldings, and BS $\sigma = 0.0075$. Each calculation scales $\sigma$ by multiplying by $S/F(S, S_0, S_\infty, x, y)$.

---

Fig. 2.

---

Fig. 3 gives an example of a two-factor distribution evolved out to $T = 0.5$ year for $x = 0.7$.

---

Fig. 3.

---

### 6.4. Two-Factor Volatility and PATHINT Modifications

In our two-factor model, the mesh of *S* would depend on $\sigma$ and cause some problems in any PATHINT grid to be developed in $S$-$\sigma$.

For some time we have considered how to handle this generic problem for *n*-factor multivariate systems with truly multivariate diffusions within the framework of PATHINT. In one case, we have taken advantage of the Riemannian invariance of the probability distribution as discussed above, to transform to a system where the diffusions have only "diagonal" multiplicative dependence [19]. However, this leads to cumbersome numerical problems with the transformed boundary conditions [20]. Another method, not yet fully tested, is to develop a tiling of diagonal meshes for each factor *i* that often are suitable for off-diagonal regions in an *n*-factor system, e.g.,

$$\Delta M_k^i = 2^{m_k^i} \Delta M_0^i \; ,$$

$$\Delta M_0^i \approx \sqrt{g^{|i||i|} \Delta t} \; , \tag{47}$$

where the mesh of variable *i* at a given point labeled by *k* is an exponentiation of 2, labeled by $m_k^i$; the integral power $m_k^i$ is determined such that it gives a good approximation to the diagonal mesh given by the one-factor PATHINT mesh conditions, in terms of some minimal mesh $\Delta M_0^i$, throughout regions of the Lagrangian giving most important contributions to the distribution as predetermined by a scan of the system. This tiling of the kernel is to be used together with interpolation of intermediate distributions.

The results of our study here are that, after the at-the-money BPV are scaled to be equivalent, there is not a very drastic change in the one-factor ATM Greeks developed below. Therefore, while we have not at all changed the functional dependence of the Lagrangian on $S$ and $\sigma$, we have determined our meshes using a diffusion for the $S$ equation as $\sigma_0 F(S, S_0, S_\infty, x, y)$, where $\sigma_0$ is determined by the same BPV-equivalent condition as imposed on the one-factor models. This seems to work very well, especially since we have taken our $\sigma$ equation to be normal with a limited range of influence in the calculations. Future work yet has to establish a more definitive distribution for $\sigma$.

## 7. CALCULATION OF DERIVATIVES

### 7.1. Primary Use of Probabilities For European Options

We can use PATHINT to develop the distribution of the option value back in time from expiration. This is the standard approach used by CRR, explicit and implicit Crank-Nicolson models, etc [56].

For European options, we also take advantage of the accuracy of PATHINT enhanced by normalizing the distribution as well as the kernel at each iteration (though in these studies this was not required after normalizing the kernel). Therefore, we have calculated our European option prices and Greeks using the most elementary and intuitive definition of the option's price $V$ [57], which is the expected value

$$V = < \max[z(S - X), 0] > , \begin{cases} z = 1 , & \text{call} \\ z = -1 , & \text{put} \end{cases} , \tag{48}$$

where $X$ is the strike price, and the expected value $< . >$ is taken with respect to the *risk-neutral* distribution of the underlying market $S$. It should be noted that, while the standard approach of developing the option price delivers at the present time a range of underlying values for a given strike, our approach delivers a more practical complete range of strikes (as many as 50–60 for Eurodollar options) for a given underlying at the present time, resulting in a greatly enhanced numerical efficiency. The risk-neutral distribution is effectively calculated taking the drift as the cost-of-carry $b$ times $S$, using the above arguments leading to the BS formula. We have designed our codes to use parameters risk-free-rate $r$ and cost-of-carry $b$ such that

$b = r$,  cost of carry on nondividend stock ,

$b = r - q$, cost of carry on stock paying dividend yield $q$ ,

$b = 0$, cost of carry on future contract ,

$b = r - r_f$, cost of carry on currency with rate $r_f$ ,                (49)

which is similar to how generalized European BS codes use $b$ and $r$ [58].

Using this approach, the European price $V_E$ is calculated as

$$V_E = < \max[z(e^{(b-r)T} S_T - e^{-rT} X), 0] > .$$                (50)

The American price $V_A$ must be calculated using a different kernel going back in time from expiration, using as "initial conditions" the option values used in the above average. This kernel is the transposed matrix used for the European case, and includes additional drift and "potential" terms due to the need to develop this back in time. This can be understood as requiring the adjoint partial differential equation or a postpoint Lagrangian in real time.

That is, a forward equation for the conditional probability distribution $P[M(t + dt), t + dt | M(t), t]$ is

$$\partial P / \partial t = \frac{1}{2} (g^{GG'} P)_{,GG'} - (g^G P)_{,G} + V ,$$                (51)

where the partial derivatives with respect to $M^G$ act on the postpoint $M^G(t + dt)$. A backward equation for the conditional probability distribution $P[M(t + dt), t + dt | M(t), t]$ is

$$\partial P / \partial t = \frac{1}{2} g^{GG'} P_{,GG'} - g^G P_{,G} + V ,$$                (52)

where the partial derivatives with respect to $M^G$ act on the prepoint $M^G(t)$. The forward equation has a particularly simple form in the mathematically equivalent prepoint path integral representation.

Above, we have described how the forward distribution at present time $T_0$ is evolved using PATHINT to the time of expiration, $P(T)$, e.g., using a path-integral kernel $K$ folded over $n$ epochs, where it is folded with the function $O$,

$$V = < O(T) P(T) > = < O(K^n P(t_0)) > ,$$

$$O(T) = \max[z(e^{(b-r)T} S_T - e^{-rT} X), 0] ,$$                (53)

to determine the European value at the present time of the calls and puts at different strike values $X$. An equivalent calculation can be performed by using the backward equation, expressed in terms of the "equivalent" kernel $K^{\dagger}$ acting on $O$,

$$V = <O(T)\,P(T)> = <(K^n O(T))P(t_0)> . \tag{54}$$

It is convenient to use the simple prepoint representation for the Lagrangian, so the backward kernel is first re-expressed as a forward kernel by bringing the diffusions and drifts inside the partial derivatives, giving a transformed adjoint kernel $K^{\dagger}$.

The above mathematics is easily tested by calculating European options going forwards and backwards. For American options, while performing the backwards calculation, at each point of mesh, the options price evolving backwards from $T$ is tested and calculated as

$$O_{\text{new}} = \max[(S - X), (e^{-rdt} O_{\text{old}})] . \tag{55}$$

The Greeks $\{\Delta, \Gamma, \Theta\}$ are directly taken off the final developed option at the present time, since the $M^G$ mesh is available for all derivatives. We get excellent results for all Greeks. Note that for CRR trees, only one point of mesh is at the present time, so $\Delta$ requires moving back one epoch and $\Gamma$ requires moving back two epochs, unless the present time is pushed back additional epochs, etc.

## 7.2. PATHINT Baselined to CRR and BS

The CRR method is a simple binomial tree which in a specific limit approaches the BS partial differential equation. It has the virtues of being fast and readily accommodates European and American calculations. However, it suffers a number of well-known numerical problems, e.g., a systematic bias in the tree approximation and an oscillatory error as a function of the number of intermediate epochs/iterations in its time mesh. Some Greeks like $\{\Delta, \Gamma, \Theta\}$ can be directly taken off the tree used for pricing with reasonable approximations (at epochs just before the actual current time). The first problem for American options can be alleviated somewhat by using the variant method [5],

$$\text{CRR}_{\text{variant}} = \text{CRR}_{\text{American}} - \text{CRR}_{\text{European}} + \text{BS} . \tag{56}$$

The second problem can be alleviated somewhat by averaging runs of $n$ iterations with runs of $n + 1$ iterations [59]. This four-fold increase of runs is rarely used, though perhaps it should be more often. Furthermore, if increased accuracy in price is needed in order to take numerical derivatives, typically 200–300 iterations should be used for expirations some months away, not 30–70 as too often used in practice.

When taking numerical derivatives there can arise a need to tune increments taken for the differentials. For some Greeks like $\Delta$ and $\Gamma$ the size of the best differentials to use may vary with strikes that have different sensitivities to parts of the underlying distribution. One method of building in some adaptive flexibility across many such strikes is to increase the order of terms used in taking numerical derivatives. (This was not required for results presented here.) For example, it is straightforward to verify that, while the central difference

$$\frac{df}{dx} = \frac{f(x+dx) - f(x-dx)}{2dx} \tag{57}$$

is typically good to $o((dx)^3)$,

$$\frac{df}{dx} = \frac{-f(x+2dx) + 8f(x+dx) - 8f(x-dx) + f(x-2dx)}{12dx} \tag{58}$$

is typically good to $o((dx)^5)$. Similarly, while

$$\frac{d^2f}{dx^2} = \frac{f(x+dx) - 2f(x) + f(x-dx)}{dx\,dx} \tag{59}$$

is typically good to $((dx)^4)$,

$$\frac{d^2f}{dx^2} = \frac{-d(x+2dx) + 16f(x+dx) - 30f(x) + 16f(x-dx) - f(x-2dx)}{dx\,dx} \tag{60}$$

is typically good to $((dx)^6)$.

Table 1 gives an example of baselining our one-factor PATHINT code to the CRR and BS codes using the above safeguards for an option whose American price is the same as its European counterpart, a typical circumstance [58]. In the literature, the CRR code is most often taken as the benchmark for American calculations. We took the number of intermediate epochs/points to be 300 for each calculation. Parameters used for this particular ATM call are $T = 0.5$ years, $r = 0.05$, $b = 0$, $\sigma = 0.10$.

________________________________________________________________________

Table 1.

________________________________________________________________________

Tests with American CRR and American PATHINT lead to results with the same degrees of accuracy.

### 7.3. Two-Factor PATHINT Baselined to One-Factor PATHINT

Previous papers and tests have demonstrated that the two-factor PATHINT code performs as expected. The code was developed with only a few lines to be changed for running any $n$-factor problem, i.e., of course after coding the Lagrangian specific to a given system. Tests were performed by combining two one-factor problems, and there is no loss of accuracy. However, here we are making some additional mesh approximations as discussed above to accommodate $\sigma$ in the $S$ diffusion. This seems quite reasonable, but there is no sure test of the accuracy. We indeed see that the ATM results are very close across $x$'s, similar to our ATM comparisons between BS and our one-factor PATHINT results for various $x$'s, where again scaling is performed to have all models used the same BPV (using the $\sigma_0$ procedure for the mesh as described above for the two-factor model).

The logical extension of Greeks for the two-factor model is to develop derivatives of price with respect to $\rho$ and $\varepsilon$ in $\sigma$ volatility equation. However, we did not find a *bona fide* two-factor proxy for the one-factor $\Upsilon$, the derivative of price with respect to the one-factor $\sigma$ constant. We get very good ATM $\Upsilon$ comparisons between BS and our one-factor models with various $x$'s. We tried simply multiplying the noise in the two-factor stochastic volatility in the price equation by a parameter with deviations from 1 to get numerical derivatives of PATHINT solutions, and this gave somewhat good agreement with the ATM BPV-scaled BS $\Upsilon$ within a couple of significant figures. Perhaps this is not too surprising, especially given the correlation substantial $\rho$ between the price and volatility equations which we do not neglect.

### 8. CONCLUSIONS

The results of our study are that, after the at-the-money basis-point volatilities are scaled to be equivalent, there is only a very small change in option prices for different exponents $x$, both for the one-factor and two-factor models. There still are significant differences in Greeks using more sophisticated models, especially for out-of-the-money options. This implies that current pricing models are not radically mispricing the markets.

Our studies point to contexts of markets well recognized by option traders to have significantly different volatility behavior. Suppression of stochastic volatility in the one-factor model just leaks out into stochasticity of parameters in the model, e.g., especially in $x$, unless additional numerical methods are employed, e.g., using an adaptive standard deviation. Our studies show that the two-factor exponents $x$ are reasonably faithful indicators defining these different contexts. The $x$-exponents in the two-factor fits are quite stable. As such, especially the two-factor $x$ can be considered as a "context indicator" over a longer time scale than other indicators typically used by traders. The two-factor fits also exhibit differences due to the $\sigma$ parameters, including the $\rho$ correlations, in accord with the sense traders have about the nature of changing volatilities across this time period.

Modern methods of developing multivariate nonlinear multiplicative Gaussian-Markovian systems are quite important, as there are many such systems and the mathematics must be diligently exercised if such models are to faithfully represent the data they describe. Similarly, sophisticated numerical techniques, e.g., global optimization and path integration are important tools to carry out the modeling and fitting to data without compromising the model, e.g., by unwarranted quasi-linear approximations. Three quite different systems have benefited from this approach:

The large-scale modeling of neocortical interactions has benefited from the use of intuitive constructs that yet are faithful to the complex algebra describing this multiple-scaled complex system. For example, canonical-momenta indicators have been successfully applied to multivariate financial markets.

It is clear that ASA optimization and PATHINT path-integral tools are very useful to develop the algebra of statistical mechanics for a large class of nonlinear stochastic systems encountered in finance. However, it also is clear that each system typically presents its own non-typical unique character and this must be included in any such analysis. A virtue of this statistical mechanics approach and these associated tools is they appear to be flexible and robust to handle quite different systems.

**ACKNOWLEDGMENTS**

I thank Donald Wilson for his support and Jennifer Wilson for collaboration running ASA and PATHINT calculations. Implied volatility and yield data was extracted from the MIM database of Logical Information Machines (LIM).

**FIGURE CAPTIONS**

FIG. 1.  (a) $F(S, S_0, S_\infty, x, y)$ for $x = 1$, the Black-Scholes case.  The other parameters are $S = 5$, $S_0 = 0.5$, $S_\infty = 20$, $y = 0$.  (b) $F(S, S_0, S_\infty, x, y)$ for $x = 0$, the normal distribution.  (c) $F(S, S_0, S_\infty, x, y)$ for $x = -1$.  (d) $F(S, S_0, S_\infty, x, y)$ for $x = 2$.

FIG. 2.  The short-time probability distribution at time $T = 0.5$ years for $x = 1$, the (truncated) Black-Scholes distribution.  The short-time probability distribution at time $T = 0.5$ years for $x = 0$, the normal distribution.  The short-time probability distribution at time $T = 0.5$ years for $x = -1$.  The short-time probability distribution at time $T = 0.5$ years for $x = 2$.

FIG. 3.  A two-factor distribution evolved out to $T = 0.5$ year for $x = 0.7$.

**TABLE CAPTIONS**

Table 1. Calculation of prices and Greeks are given for closed form BS (only valid for European options), binomial tree $CRR_{European}$, $CRR_{American}$, $CRR_{variant}$, and PATHINT. As verified by calculation, the American option would not be exercised early, so the PATHINT results are identical to the European option. The $CRR_{American}$ differs somewhat from the $CRR_{European}$ due to the discrete nature of the calculation. All CRR calculations include averaging over 300 and 301 iterations to minimize oscillatory errors.

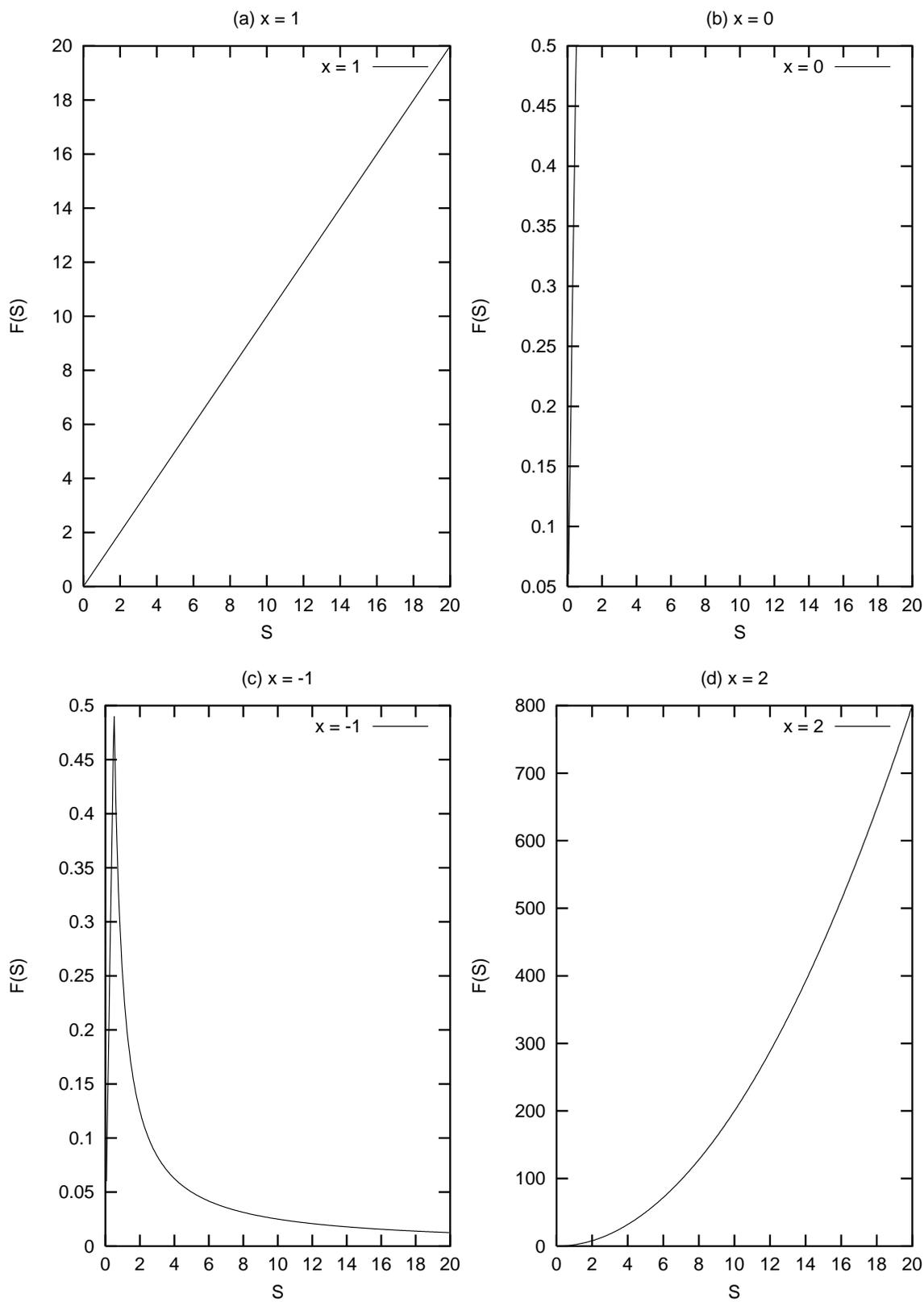

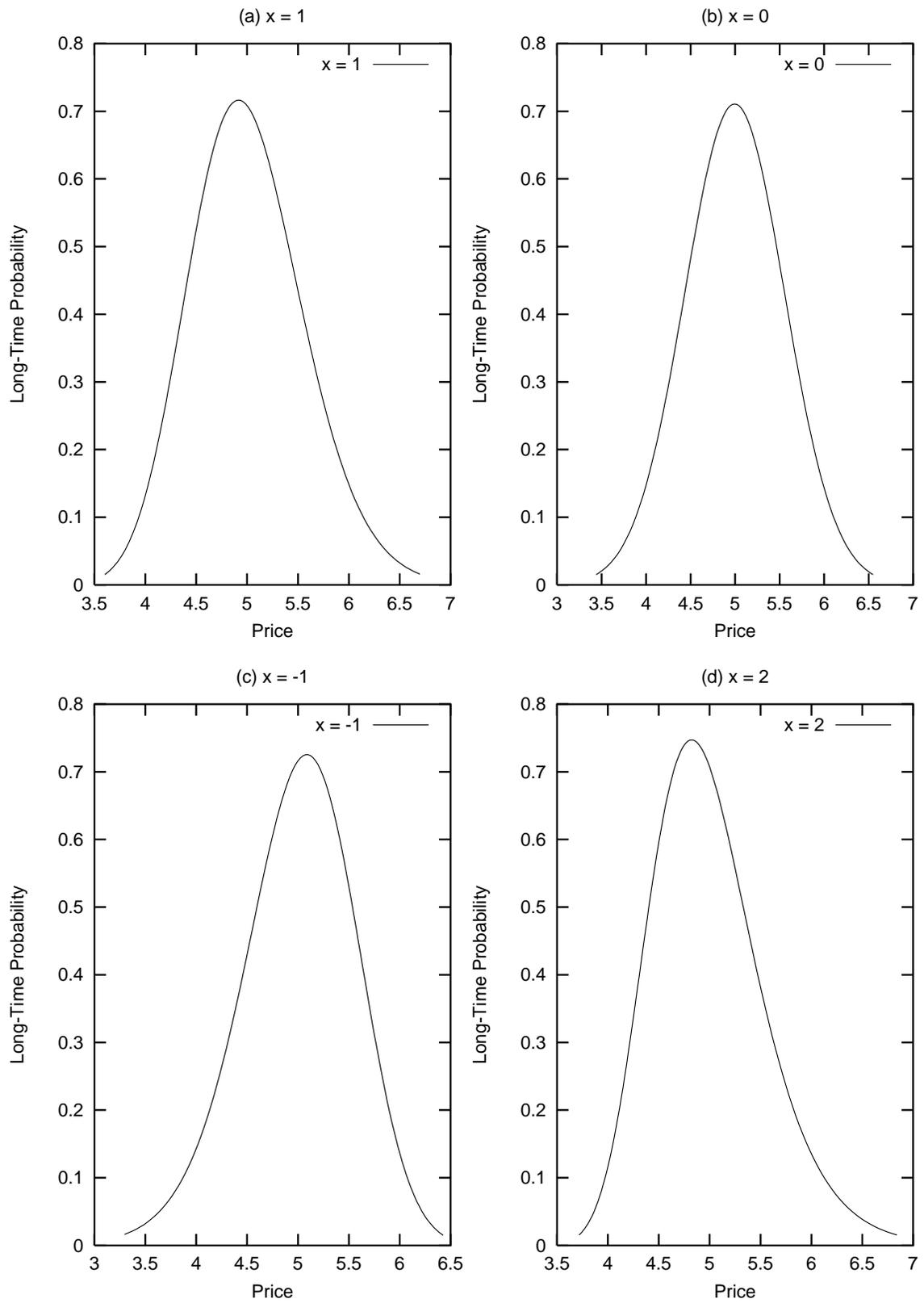

Two-Factor Probability ———

Long-Time Probability

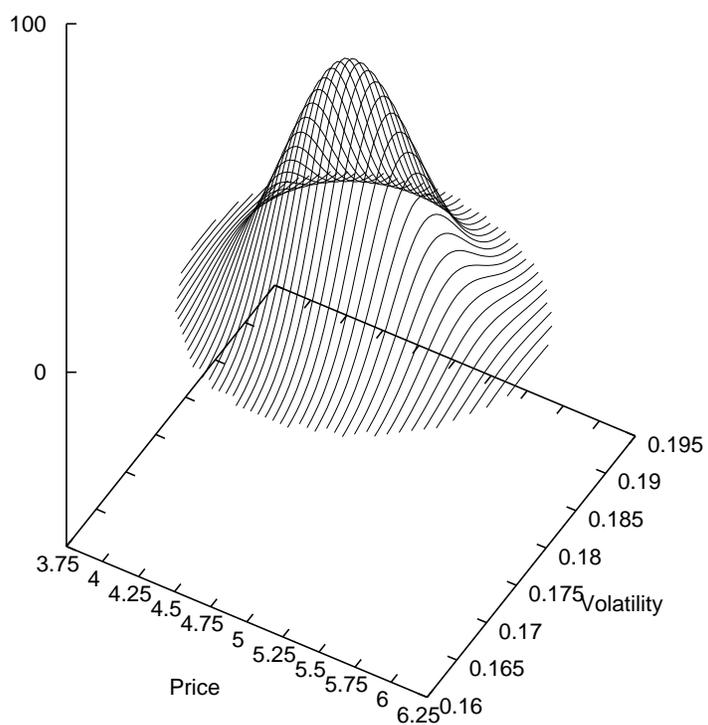

| Greek | BS | CRR$_{European}$ | CRR$_{American}$ | CRR$_{variant}$ | PATHINT |
|-------|------|------|------|------|------|
| Price | 0.138 | 0.138 | 0.138 | 0.138 | 0.138 |
| Delta | 0.501 | 0.530 | 0.534 | 0.506 | 0.501 |
| Gamma | 1.100 | 1.142 | 1.159 | 1.116 | 1.100 |
| Theta | -0.131 | -0.130 | -0.132 | -0.133 | -0.131 |
| Rho | -0.0688 | -0.0688 | -0.0530 | -0.0530 | -0.0688 |
| Vega | 1.375 | 1.375 | 1.382 | 1.382 | 1.375 |